\documentclass{emulateapj}
\usepackage{times}

\shorttitle{Continuum Acceleration of Black Hole Winds}
\shortauthors{Everett \& Ballantyne}

\begin{document}

\title{Continuum Acceleration of Black Hole Winds} 

\author{John E. Everett and David R. Ballantyne} 
\affil{Canadian Institute for Theoretical Astrophysics, University of
Toronto, 60 St. George Street, Toronto, ON M5S 3H8, Canada}
\email{everett, ballantyne@cita.utoronto.ca} \slugcomment{Accepted to ApJL}

\begin{abstract}
Motivated by recent observations of high-velocity, highly ionized
winds in several QSOs, models of purely continuum-driven winds
launched from $\sim200~GM_{\rm BH}/c^2$ are presented.  Launching
conditions are investigated, as well as the observational signatures
for a variety of initial conditions and illuminating continua.  While
we verify that continuum-driven, highly-ionized outflows reach the
observed velocities for $L/L_{\rm Edd}\geq1$, independent of the
incident spectral shape, such winds are too highly ionized to exhibit
the observed absorption features when launched with an AGN continuum
(in fact, such winds are so ionized that they are driven primarily by
electron scattering). If the wind is instead illuminated with a
blackbody continuum originating from an optically thick shield, the
gas is too weakly ionized and does not produce high energy absorption
features.  If high-velocity, high-ionization winds are truly launched
from very near the black hole, such winds must be launched under other
conditions or via other processes; we summarize some possibilities.
\end{abstract}


\section{Introduction}\label{Intro}

Optical, UV, and X-ray absorption troughs in Active Galactic Nuclei
(AGN) spectra are central to the ongoing effort to understand not only
mass outflow in AGN, but to provide clues to the fueling of their
central supermassive black holes \citep[see][]{CKG03}.  Although the
optical depth, metallicity, and velocity of such outflows are becoming
increasingly well-constrained, the mechanism accelerating the winds
remains elusive: thermal \citep[e.g.,][]{KK95}, continuum-driven
\citep[for instance,][]{CN03}, line-driven
\citep[e.g.,][]{MCGV95,PSK00}, and hydromagnetic \citep[e.g.,][]{KK94,
B97} winds may all provide insight, but none of the above models have
been excluded or are definitively observationally preferred for AGN
outflows.

In this Letter, observational constraints on continuum-driven winds
launched from very near the central black hole are examined.  Such
wind models have become significant, as observations of outflows
apparently moving at large fractions of the speed of light
($v~\sim~$0.1--0.4c) have been published by \citet{CBG03},
\citet{CBGG03}, \citet{Pounds03a, Pounds03b}, and \citet{ROW03}. From
the measured high velocities, the winds are inferred to be launched
very close to the central black hole.  \citet{KP03} have proposed that
the highly-ionized gas is subject to continuum driving near the
central black hole because line driving is unlikely to be important
for the large ionization parameters in these winds.

A model of a continuum-driven wind is presented in
\S\ref{modelOverview}. This model is used to investigate terminal
velocities and absorption features displayed by a variety of winds
with different incident continua, initial shielding columns, and
densities at the base of the wind (\S\ref{Results}).  


\section{Model Overview}\label{modelOverview}

Our simple continuum-driven wind model consists of a radial streamline
set to rise above the disk at a constant angle of $5^\circ$ to the
disk, with the initial column ($N_{\rm H,0}$) and number density
($n_0$) of the wind specified by the user. Gas with that initial
density is assumed to be in Keplerian motion and is perturbed into the
central, illuminating continuum with an initial velocity of roughly
the local sound speed.  To determine the gas's motion due to continuum
acceleration, Cloudy \citep{Ferland02} simulations are run on a
radial rays through the wind, spaced logarithmically in co-latitude.
The continuum acceleration on the gas is then computed using both the
ionization state and transmitted continuum (therefore taking account
of any self-shielding effects) at the end of each radial slice through
the wind, taking into account both electron scattering and bound-free
opacities.
The radiative acceleration is calculated and then compared to gravity
in $\Gamma_{\rm cont}$, defined as:
\begin{equation}
\Gamma_{\rm cont} = \frac{a_{\rm cont}}{g} = \frac{\frac{1}{\rho c} \int \chi_{\nu} F_{\nu}
  d\nu}{\frac{G M_{\rm BH}}{R^2}}
\end{equation}
where $a_{\rm cont}$ is the continuum radiative acceleration, $g$ is
the gravitational acceleration, $\rho$ is the density in the wind, $c$
is the speed of light, $\chi_{\nu}$ is the frequency-dependent
continuum opacity (output by Cloudy), $F_{\nu}$ is the transmitted
continuum (again, given by Cloudy), $G$ is the gravitational constant,
$M_{\rm BH}$ is the mass of the central black hole, and $R$ is the
spherical radius.  This $\Gamma_{\rm cont}$ factor is then input into
the Euler equation along with the component of gravity and centrifugal
acceleration along the radial flowline:
\begin{equation}
v\frac{dv}{dr}=-\frac{G M_{\rm BH}(1-\Gamma_{\rm cont})}{R^2}
           \cos(\theta-\theta_{\rm F}) + 
           \frac{v_{\phi}^2(\varpi)}{\varpi}\sin\theta_{\rm F}
\end{equation}
where $\varpi$ is the cylindrical radius, $\theta$ is the co-latitude,
$\theta_{\rm F}$ is the (fixed) angle of the flowline from the
vertical, and $v_{\phi}(\varpi) = \Omega_0 r_0^2/\varpi$ is the
azimuthal velocity (angular momentum is conserved in outflowing gas).

We integrate this equation to compute the velocity of the wind as a
function of distance along the radial streamline.  As that velocity
increases, the number density in the wind (as well as the column
density) decreases via mass conservation.  For the first set of Cloudy
simulations, where no velocity data are available, the density is set
to fall as $r^{-2}$; this initial density distribution is replaced in
subsequent iterations and is unimportant for the final results.  The
solution is iterated at least twice to ensure that the density
structure is consistent with the velocity profile in the wind;
simulations in test cases have shown that the wind structure does not
change after two iterations, even when ten iterations are completed.

The wind parameters chosen in this paper reproduce the observational
constraints on the outflow observed by \citet{Pounds03a} in the narrow
emission line quasar PG~1211+143.  They observed an outflow with
$v\sim0.08c$ and an inferred density of $n\sim3\times10^8~{\rm
cm}^{-3}$, situated at $3\times10^{15}~{\rm cm}$ from the black hole
\citep[this density is estimated observationally from the emission
measure and line-of-sight column density; see][]{Pounds03a}.  They
also inferred a launching radius of $r_0=10^{15}~{\rm cm}$ by noting
that $v=0.08c\sim v_{\rm esc}$, the escape velocity for that $r_0$.
Since $\dot{M}=4 \pi b r^2 \rho v$ (where $b$ is the covering factor
of the flow, $r$ is the radial distance, $\rho$ is the mass density,
and $v$ the velocity of the wind) must be constant along the flow, an
original density of $\sim5\times10^{12}~{\rm cm}^{-3}$ is inferred for
the base of the wind if the initial velocity is of order $20~{\rm
km/s}$ and $b=0.8$ \citep[the value of $b$ used in][]{Pounds03a}.
Observations indicate that $M_{\rm BH}\approx4\times10^{7}~M_{\sun}$
\citep{Kaspi00}\footnote{A recent reanalysis by \citet{Peterson04}
yields $1.46\pm0.44\times10^8~M_{\sun}$ (with significant uncertainty
noted in the analysis); we retain the value of
$4\times10^{7}~M_{\sun}$ to compare with previous models.}, and
$L_{\rm Edd}\approx5\times10^{45}~{\rm erg/s}$ \citep[$L/L_{\rm Edd}
\sim 0.8$ as the ``Big Blue Bump'' that dominates the continuum in
PG~1211+143 has $L\sim 4\times10^{45}$~erg/s;][]{Pounds03a}.  The
inferred mass outflow rate from the observations is $\sim
2.8~M_{\sun}/{\rm yr}$ (for the covering fraction $b=0.8$). We assume
solar abundances.

The increase of density towards the base of the wind (where the gas is
outflowing much more slowly) results in a large increase in the column
density over the observed column of $\sim5\times10^{23}~{\rm
cm}^{-2}$.  As recognized by \citet{KP03}, this column would be
significantly optically thick, precluding photoionization simulations.
Two possibilities are therefore considered in our simulations.  First,
``standard'' winds are tested with $N_{\rm H,0} < 10^{24}~{\rm
cm}^{-2}$ at the base of the outflow and with an AGN spectral energy
distribution (SED).  Next, winds are launched by a soft X-ray/UV
blackbody from the photosphere of an optically thick shield, as
hypothesized by \citet{KP03}.

\section{Kinematics and Absorption Features}\label{Results}

\begin{figure}[ht]
\begin{center}
\plotone{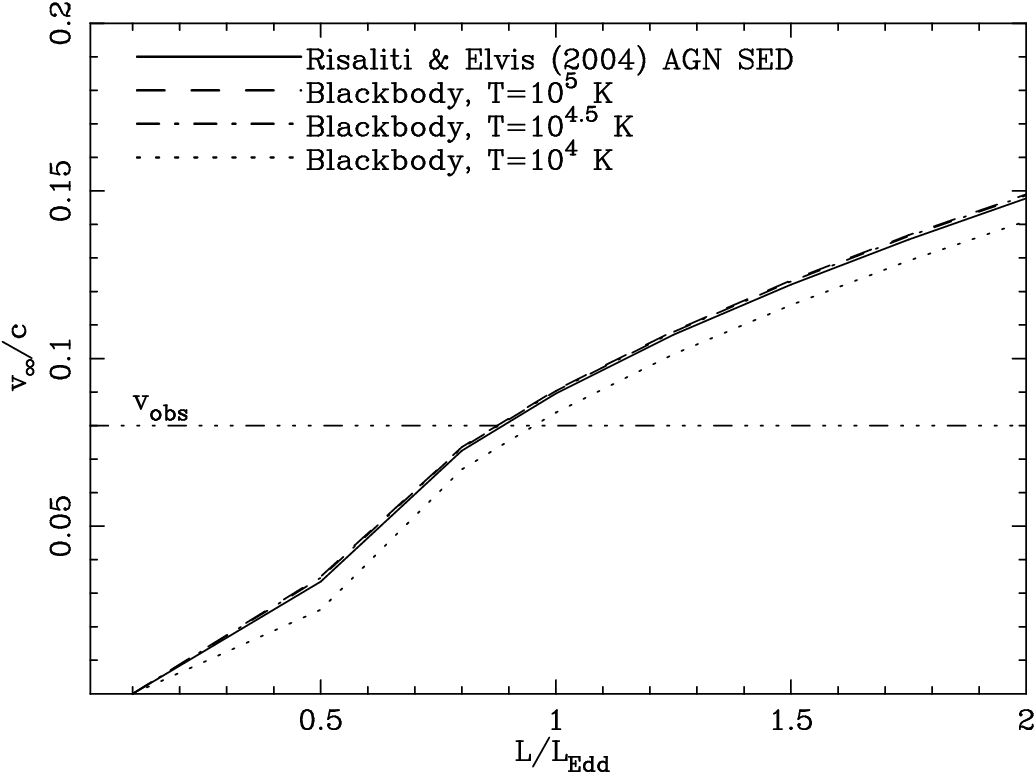}
\caption{The final velocity of continuum-driven wind models as a
function of the Eddington ratio of the central continuum for a range
of incident SEDs.  For continuum driving in such highly ionized
sources, the final velocity is relatively independent of the SED.  The
observed velocity of $v_{\rm obs}=0.08c$ in PG~1211+143 is shown for
comparison.  Note that, for $L/L_{\rm Edd} \la 1$, winds do not
escape the system, but we plot their velocities at $s=100r_0$ to show
the trend as $L/L_{\rm Edd}$ increases.
\label{vInfPlot}}
\end{center}
\end{figure}

The most immediate check for these outflows is to see which parameters
lead to the observed velocities.  Figure~\ref{vInfPlot} displays
$v_{\infty}$ as a function of the Eddington ratio ($v_{\infty}$ here
is the velocity at the end of the model's streamline at a distance of
$100r_{0}$ where $r_0$ is the launch radius of our wind; taking the
terminal velocity at $10^4~r_0$ only changes $v_{\infty}$ by $\sim1\%$
in tests).  This figure shows the variation in $v_{\infty}$ for both
an AGN SED \citep{RE04}\footnote{The \citet{RE04} spectrum is
approximated by using Cloudy's ``AGN'' continuum with the Big Blue
Bump peaking at $T=1.5\times10^5$~K, $\alpha_{\rm OX}=-1.43$,
$\alpha_{\rm UV}=-0.44$, $\alpha_{\rm X}=-0.9$.} and for a blackbody
continuum with a range of temperatures \citep[as proposed by][]{KP03}.
As a simple starting case, the winds presented here are optically
thin, with $N_{\rm H,0}=10^{22}~{\rm cm}^{-2}$; higher columns are
considered below.

As expected, when $L/L_{\rm Edd}~<~1$, the wind does not reach $v_{\rm
obs} \sim v_{\rm esc}$, even with the added bound-free opacity used to
calculating the continuum acceleration.  For $L/L_{\rm Edd}~\geq~1$,
winds can achieve the observed velocities.  Figure~\ref{vInfPlot}
clearly shows that the blackbody continuum can also accelerate a wind,
resulting in almost identical velocities as when the \citet{RE04} SED
is used.  This seems reasonable since the majority of $\Gamma_{\rm
cont}$ is given by frequency-independent electron scattering.

\begin{figure}[h]
\begin{center}
\plotone{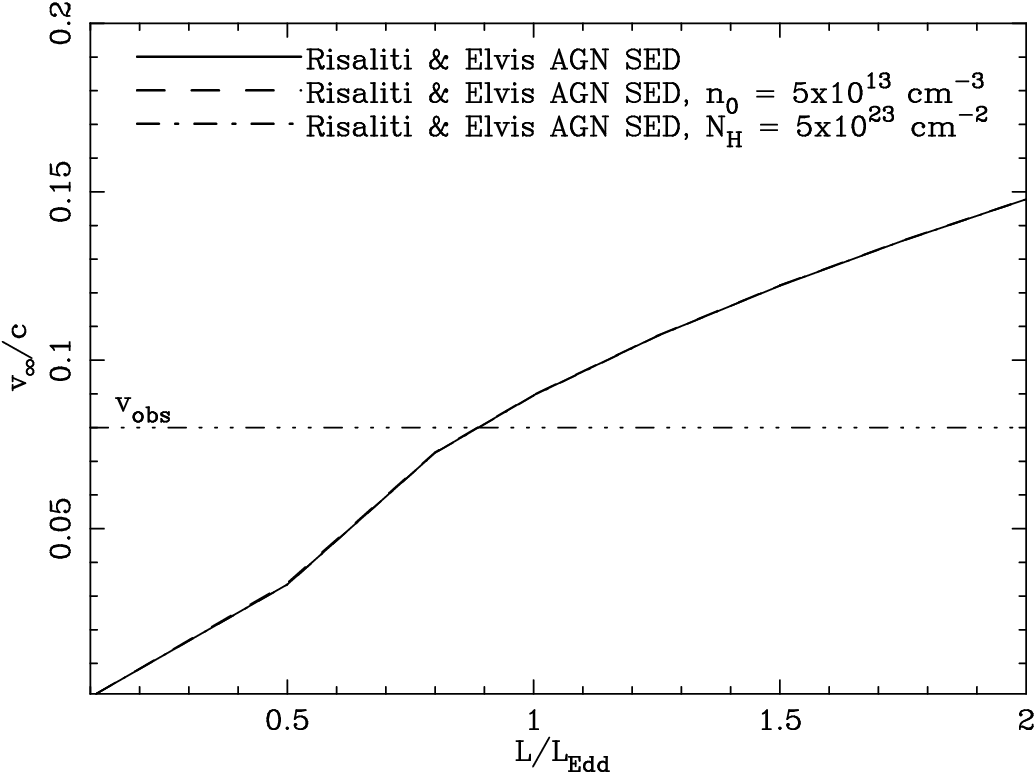}
\caption{The final velocity of continuum-driven wind models
  vs. $L/L_{\rm Edd}$ for the \citet{RE04} incident spectrum with
  changes to the initial density and initial wind column.  Unless
  otherwise noted in the legend, the models have $N_{\rm
  H,0}=10^{22}~{\rm cm}^{-2}$ and $n_0=5\times10^{12}~{\rm cm}^{-3}$.
  Neither the increased initial column nor density affect the wind
  significantly.  The observed velocity of $v_{\rm obs}=0.08c$ in
  PG~1211+143 is shown for comparison.
\label{vInfPlotRE04}}
\end{center}
\end{figure}

Next, to test the sensitivity of these results to initial parameters,
both the column and the initial density of the wind are varied (see
Fig.~\ref{vInfPlotRE04}) for winds illuminated with the AGN SED from
\citet{RE04}.  This shows that changing the initial density by an
order of magnitude (which increases $\dot{M}$ by the same factor) does
not change the velocity structure of the wind significantly, due to
the fact that the continuum is strong enough to highly ionize both
winds. For such highly-ionized continuum driven winds, the degree of
acceleration is governed by $n_{\rm e}/\rho$ which is nearly identical
for both winds (although the higher-density wind does have a lower
ionization parameter near the base of the wind, yielding larger
accelerations there than in the lower-density wind).

Also in Figure~\ref{vInfPlotRE04}, the shielding column at the base of
the wind is varied from the normal $N_{\rm H,0}=10^{22}~{\rm cm}^{-2}$
to $N_{\rm H,0}=5\times10^{23}~{\rm cm}^{-2}$ to investigate the
``self-shielding'' of a thicker wind.  This thicker column also does
not change the final velocities in the outflow.

\begin{figure}[ht]
\begin{center}
\plotone{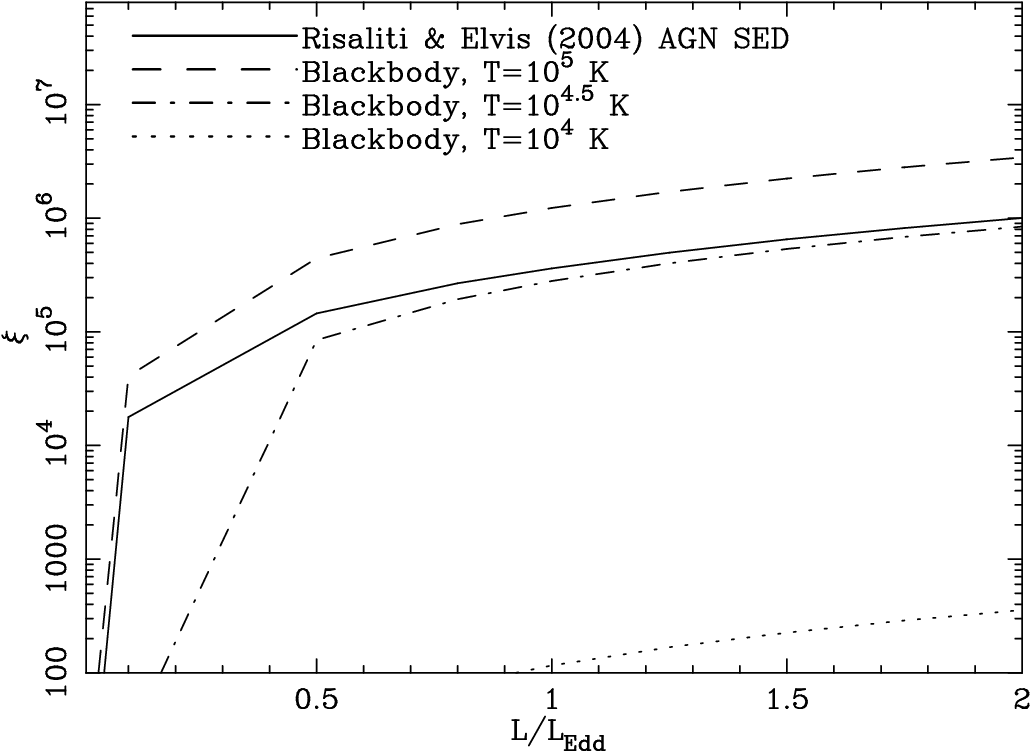}
\figcaption{The X-ray ionization parameter, $\xi$, as a function of
$L/L_{\rm Edd}$ for a range of incident continua at the inferred
location of the absorbing gas in PG~1211.  The relative values of the
ionization parameter are set by the amount of flux each SED has in the
range of 1 to 1000 Rydbergs, where $\xi$ is calculated.  The AGN SED
yields ionization parameters too high to explain the observed line
absorption.  While the blackbody continua can have relatively high
ionization parameters, they have no flux in the X-ray regime to ionize
iron to the states observed in PG~1211+143.
\label{xiPlot}}
\end{center}
\end{figure}

\begin{figure}[h]
\begin{center}
\plotone{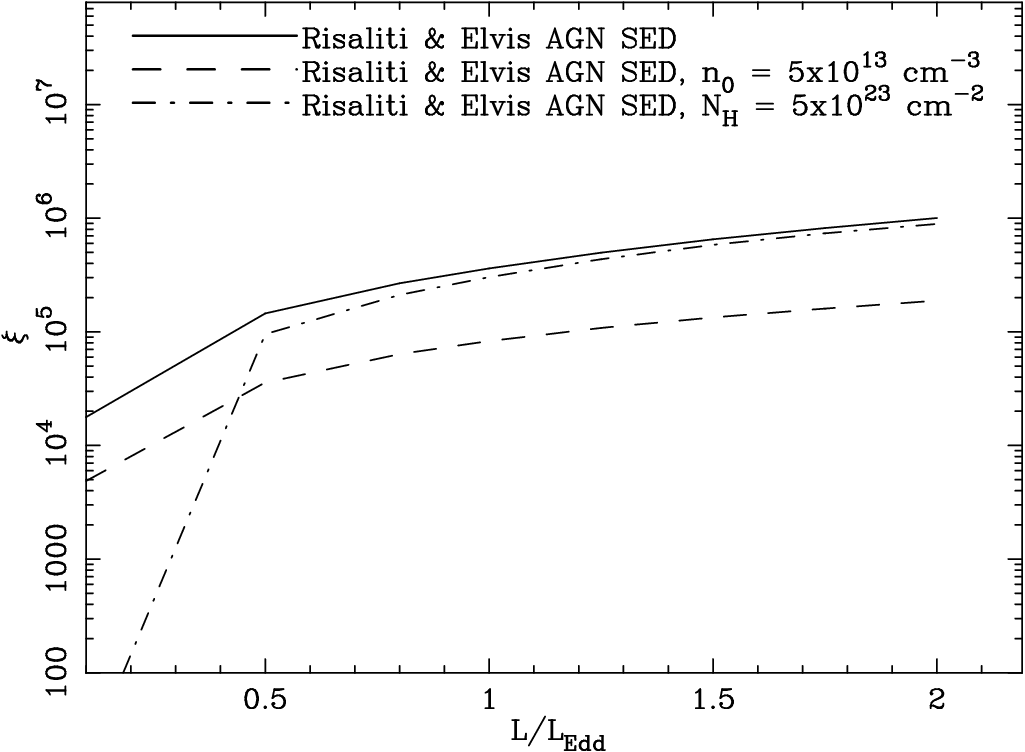}
\figcaption{As in Fig.~\ref{xiPlot}, but using the \citet{RE04}
continuum to investigate variations in density and shielding.  Unless
otherwise noted in the legend, the models have $N_{\rm
H,0}=10^{22}~{\rm cm}^{-2}$ and $n_0=5\times10^{12}~{\rm cm}^{-3}$.
The larger density wind yields a somewhat smaller $\xi$, even though
it experiences a very similar acceleration to the lower-density model.
The $N_{\rm H,0}=5\times10^{23}~{\rm cm}^{-2}$ wind has a very similar
ionization parameter for $L/L_{\rm Edd}\geq1$.
\label{xiPlotRE04}}
\end{center}
\end{figure}

These figures address the range of Eddington ratios necessary to
launch a wind to the observed velocity.  But what are the particular
spectral signatures of such a wind?  To answer this question,
Figures~\ref{xiPlot} and \ref{xiPlotRE04} show the X-ray ionization
parameter, $\xi$, as a function of $L/L_{\rm Edd}$ for those different
models.  We define $\xi$ in the same manner as Cloudy and XSTAR: $4
\pi F_{\rm X}/n_{\rm H}$, where $F_{\rm X}$ is the integrated X-ray
flux from 1 to 1000 Rydbergs.  To pick one `fiducial' $\xi$ out of the
range of ionization parameters in the accelerating wind, $\xi$ is
extracted at a distance along the flow of $s=3r_0$ (chosen to compare
with the inferred distance of the wind in PG~1211+143).

Figures~\ref{xiPlot} and \ref{xiPlotRE04} display the ionization
parameters in these continuum-driven winds.  They show that the high
velocity, continuum-driven winds must by necessity have been
accelerated by a large incident flux, which, for AGN-like continua,
leads to extremely high ionization parameters in the wind ($\xi$ of
order $10^5-10^6$ for $L/L_{\rm Edd}\geq1$).  The $T=10^5~$K blackbody
spectrum also has a very high ionization parameter, but all of its
energy in the 1-1000 Rydberg window is concentrated near 1 Rydberg.
Thus, even though this continuum has a large ionization parameter, the
lack of X-ray photons means that the ionization stages within the gas
are much lower than that produced by an AGN SED: for instance, the
predominant state of iron for the wind illuminated by the $T=10^5~$K
blackbody spectrum is Fe~\textsc{ix}, while the observed iron ions are
Fe~\textsc{xxv} and Fe~\textsc{xxvi} \citep{Pounds03a}.

This has important implications for observations of such winds.
\citet{KP03} postulated that continuum driven winds could account for
high-velocity X-ray absorption features.  As mentioned in the
introduction, there have been other recent observations of such
high-velocity X-ray absorption in a variety of QSOs \citep{CBG03,
CBGG03, Pounds03b, ROW03}.  In observations of PG~1211+143,
significant equivalent widths (EWs) for absorption features in
Mg~\textsc{xii}, S~\textsc{xvi}, Fe~\textsc{xxv}, and Fe~\textsc{xxvi}
are seen.  Table~\ref{ewTable} compares these observations to models
of a high-column wind with an AGN SED \citep{RE04}, $N_{\rm H,0} =
5\times10^{23}~{\rm cm}^{-2}$, assumed line widths of 1000 km/s
\citep{Pounds03a} and $L/L_{\rm Edd}~=~1$.  Equivalent widths are
calculated in the linear weak line limit:
\begin{equation}
W_{\rm eV}=\Delta E_{\rm D}
 \sqrt{\pi}\tau_0\times[1-\tau_0/(2!\sqrt{2})+\tau_0^2/(3!\sqrt{3})+...]
\end{equation}
for $\tau_0 \la 1$, where $\tau_0$ is the optical depth in the line
from Cloudy and $\Delta E_{\rm D}$ is the Doppler width of the line in
eV (all of the lines shown here have $\tau_0 \la 1$ in the wind).
A comparison of observed EWs to the model EWs clearly shows that,
because of the high ionization in the outflow, such winds cannot
account for the EWs observed in PG~1211+143.  The EWs for winds
illuminated with a blackbody spectrum are even smaller, although in
that case the small EWs are due to the lack of high-energy ionizing
X-rays.

To test the particular parameters presented by \citet{Pounds03a}, we
assembled a model with a constant column of $N_{\rm
H}=5\times10^{23}~{\rm cm}^{-2}$ present throughout the wind.
Although the density still declines with height above the disk as
before, we set the column to a constant value for every line of sight
through the wind.  This would therefore yield the column of $N_{\rm
H}=5\times10^{23}~{\rm cm}^{-2}$ at $s=3r_0$ as observed by
\citet{Pounds03a}.  This model differs strongly from the other
solutions where the column drops consistently as the density in the
wind falls.  The last column of Table~\ref{ewTable} displays the EWs
from this model with $L/L_{\rm Edd}=1$; at higher $L/L_{\rm Edd}$, the
EWs only decrease. Even with the constant column supplying a greater
column of the relevant ions, the ionization parameter in the model is
so high that the line equivalent widths fall below the 90\% confidence
interval of the measurements.

We have searched for some area of parameter space that might allow
such continuum-driven winds to show the observed equivalent widths.
The only possibility we have found is modifying the incident continuum
such that $\alpha_{\rm x}=-2$ ($\Gamma = 3$).  While inconsistent with
the observed power law of $\Gamma \sim 1.75$, this does yield
equivalent widths consistent with the observations for all lines
except Fe~\textsc{xxv}, but only for the wind with constant column of
$N_{\rm H} = 5\times10^{23}~$cm$^{-2}$ along the entire outflow.
However, in the rather implausible case that the X-ray continuum is
formed outside the wind (so that the wind sees a different X-ray
continuum than is observed), such a model does reproduce the observed
equivalent widths.

As a final consideration, we note that for a large outflow with
covering factor $b=0.8$, P-Cygni profiles should be observed, but
emission is not seen with the observed absorption lines.  This may
perhaps indicate a smaller covering fraction for the wind.  

\begin{deluxetable}{lccc}
\tabletypesize{\scriptsize}
\tablecaption{Comparison of equivalent widths for PG~1211+143
  \citep[from][]{Pounds03a} to a continuum driven
  wind.\label{ewTable}} 
\tablehead{
\colhead{Line} & \colhead{Observed} &
  \colhead{Model EW [eV]\tablenotemark{b}} & 
  \colhead{Model EW [eV]\tablenotemark{b}\tablenotemark{c}} \\
 & \colhead{EW [eV]\tablenotemark{a}} & 
  \colhead{$N_{\rm H,0}=5\times10^{23}~{\rm cm}^{-2}$} &
  \colhead{$N_{\rm H}=5\times10^{23}~{\rm cm}^{-2}$}   
}
\startdata
Fe~\textsc{xxvi} Ly$\alpha$   &   $95 \pm 20$     &  $7\times10^{-4}$  &  31 \\
Fe~\textsc{xxv} 1s-3p         &   $45 \pm 12$     &  $2\times10^{-5}$  &  0.2 \\
Fe~\textsc{xxvi} Ly$\beta$    &   $45 \pm 12$     &  $1\times10^{-3}$  &  8 \\
S~\textsc{xvi} Ly$\alpha$     &   $32 \pm 12$     &  $2\times10^{-4}$  &  1.4 \\
Mg~\textsc{xii} Ly$\alpha$    &   $15 \pm 6 $     &  $6\times10^{-5}$  &  0.4 \\
\enddata

\tablenotetext{a}{Errors listed correspond to 90\% confidence
intervals.}  
\tablenotetext{b}{Equivalent widths are those found at $s=3r_0$, the
inferred location of the sight-line through the outflow in PG~1211+143.}
\tablenotetext{c}{The column for this calculation is held constant
  throughout the outflow, unlike all previous models, to compare with
  \citet{Pounds03a}.}

\end{deluxetable}

\section{Summary}\label{Summary}

This Letter addresses observational constraints on winds launched from
near the central black hole of AGNs using pure continuum driving.
This has become an increasingly significant topic very recently, with
the observation of high-velocity, very highly ionized gas in several
different QSOs, where the outflow is most likely too highly ionized to
allow for significant line driving.  As expected, continuum driving
can accelerate highly-ionized winds from near the central black hole
to the observed velocities when $L/L_{\rm Edd} \geq 1$.  But the
models also indicate that:

\begin{enumerate}
\item Changes in the initial density or shielding column do not affect
  the final velocity of the wind.  Nor do changes in the illuminating
  continuum significantly change the final velocities in the wind,
  although moving to softer continua does significantly lower the
  ionization state.
\item Continuum-driven winds, illuminated by an AGN SED and
  accelerating to the observed velocities quickly become very highly
  ionized (with $\xi\sim 10^5-10^6$ for $L/L_{\rm Edd} \geq 1$),
  completely stripping the outflowing gas, resulting in acceleration
  primarily by electron scattering.  Continuum-driven winds can be
  launched by blackbody SEDs but such outflows do not have the
  observed high-ionization lines.
\item Taking the particular example of the outflow observed in
  PG~1211+143 \citep{Pounds03a}, wind models display much smaller EWs
  than the observed EWs.  Softening the X-ray power law to $\Gamma
  \sim 3$ can increase the EWs, however this requires both a
  non-standard AGN continuum and a constant-column disk wind.
\end{enumerate}

If pure continuum driving has difficulty reproducing the observations,
what process is accelerating the wind?  First, perhaps the wind is not
relativistic \citep[see][who finds the low-energy X-ray spectrum of
PG~1211+143 to be consistent with a 3000 km/s outflow]{Kaspi04}.  It
may also be possible that line-driving is playing a role, although
such acceleration would have to occur only at the very base of the
wind, where the ionization parameter is suitably low and where
radiative launching would be difficult given the high column to the
central source near the disk \citep[but see][]{PK04}.  Another option
for explaining these outflows is a multi-phase wind, although such a
wind would increase the mass outflow rate beyond its already large
value.

Another possibility is that such high-velocity, highly-ionized winds
could be launched hydromagnetically.  The addition of hydromagnetic
driving would decouple the ionization state from the acceleration
process; instead of requiring, as for continuum-driven flows, that the
luminosity be $L_{\rm Edd}$ or greater (which leads to
over-ionization), the wind could possibly be accelerated by MHD forces
that would allow a lower $L/L_{\rm Edd}$ and a lower ionization state.
Indeed, a lower $L/L_{\rm Edd}$ may already be indicated by the larger
$M_{\rm BH}$ (albeit with large uncertainty) found by
\citet{Peterson04}.  MHD wind models will be addressed in a subsequent
paper, but we point out that the large mass outflow rates ($\ga
\dot{M}_{\rm Edd}$) inferred in this source may be too high for an MHD
outflow.  Model predictions for winds driven by all of the above
processes will be useful in determining which forces are capable of
launching these outflows and which model outflows are consistent with
the observational signatures.

\acknowledgments 

We thank the referee for helpful comments that improved the paper, and
Doron Chelouche, Arieh K\"onigl, \& Norm Murray for their comments.
This work is supported by the Natural Sciences and Engineering
Research Council of Canada.

\end{document}